\documentclass[twocolumn,showpacs,amsmath,amssymb,superscriptaddress]{revtex4}
\usepackage{graphicx,amsmath,amssymb,amsfonts,sidecap,color}

\newcommand{\sub}[1]{_{\rm #1}}

\newcommand{\vctr}[1]{{\boldsymbol {#1}}}

\newcommand{\ket}[1]{\ensuremath{\left|#1\right\rangle}}

\newcommand{\ve}{\mathbf}

\begin{document}

\title{Exchange-based CNOT gates for singlet-triplet qubits  with   spin orbit interaction}

\author{Jelena Klinovaja}
\affiliation{Department of Physics, University of Basel,
             Klingelbergstrasse 82, CH-4056 Basel, Switzerland}
\author{Dimitrije Stepanenko}
\affiliation{Department of Physics, University of Basel,
             Klingelbergstrasse 82, CH-4056 Basel, Switzerland}
\author{Bertrand I.~Halperin}
\affiliation{Department of Physics, Harvard University, 17 Oxford St., 5 Cambridge, MA 02138, USA}
\author{Daniel Loss}
\affiliation{Department of Physics, University of Basel,
             Klingelbergstrasse 82, CH-4056 Basel, Switzerland}

\date{\today}

\begin{abstract}
   
We propose a scheme for implementing the CNOT gate over
qubits encoded in a pair of electron spins in a double quantum dot.
The scheme is based on exchange and spin orbit interactions and
on local gradients in Zeeman fields. We find that the optimal 
device geometry for this implementation involves 
effective magnetic fields that are parallel to the
symmetry axis of the spin orbit interaction.  
We show that the switching times
for the CNOT gate can be as fast as a few nanoseconds
 for realistic parameter values in GaAs semiconductors.
Guided by recent
advances in surface codes, we also consider the perpendicular
geometry.  In this case, leakage errors due to spin orbit
interaction occur but can be suppressed in  strong magnetic fields.
\end{abstract}

\maketitle


\section{Introduction}

Standard quantum computing \cite{NC00} is based on encoding,
manipulating, and measuring quantum information encoded in the state of
a collection of quantum two-level systems - qubits. Spin-$1/2$ is an
ideal implementation of a qubit, since it is a natural two-level
system, and every pure state of a spin-$1/2$ corresponds to a state of
a qubit.  For this reason, spins have been considered as carriers of
quantum information in a variety of proposals \cite{KL12}.  The
initial proposal \cite{LD98} called for spins in single-electron
quantum dots electrically manipulated by the exchange interaction and local
time-dependent Zeeman fields.  A variety of other encoding schemes
and manipulation techniques \cite{L02,B01,DBK+00,PJT+05,TGL07,TGL08,Yacoby_ST,TDT+12}
rely upon encoded qubits.  In these schemes, the simplicity of qubit
states and minimal number of physical carriers of quantum information
are traded for less stringent requirements for experimental
implementations.  On one hand, the alternative setups  protect from
the most common types of errors by decoupling the computational
degrees of freedom from the most common sources of noise, and
therefore allow for longer gating times.  On the other hand, in some
alternative setups the manipulation without fast switching of the
local magnetic fields becomes possible.

The optimization in the encoding and manipulation protocols is always
guided by the state of the art in the experiments.  Recent results
suggest that spin qubits can reside in a variety of material hosts
with novel properties.  Quantum dots in graphene
\cite{TBL+07} and carbon nanotubes
\cite{BTL08} are less susceptible to the decoherence
due to nuclei and spin-orbit interaction.  Spins in nanowires show
very strong confinement in two spatial directions, and the gating is
comparably simple \cite{TGL07,TGL08}.  In hole systems, the
carriers have distinct symmetry properties, and show coupling to the
nuclear spins of a novel kind \cite{FCB+08}.  Recently, the
experiments in silicon \cite{MBH+12}  have demonstrated coherent
manipulation of spins similar to the one achieved in the GaAs-based
nanostructures.  Within these hosts, the manipulation techniques that use exchange interaction, spatially
inhomogeneous time independent Zeeman splitting, and nuclear
hyperfine interactions are within reach.  Despite these developments, GaAS remains a promising route
to spin qubits due to the highly advanced experimental techniques developed for this material.

Here, we study the
implementation of the quantum gates on the encoded two-spin singlet-triplet (ST) qubits \cite{B01,L02,PJT+05,JPM+05,BFN+11,Yacoby_ST} using resources
that closely resemble the ones available in the current experimental
setups.  There, the application of the time-dependent electric fields
through the gates fabricated into a structure \cite{TDT+12}
are preferred to  time-dependent local magnetic fields.  In
addition, the nuclear spins \cite{PJT+05,JPM+05,TKK+06,RPB10,BFN+11} and
inhomogeneous magnetic fields \cite{BKO+07,BSO+11} are possible
resources for spin control.  In the setups based on semiconductors,
the electrons or holes in the quantum dots are influenced by the
spin orbit interaction (SOI), which can contribute to the control \cite{SRH+12}.

In this work we present a scheme for control of ST-qubits which uses
switching of the exchange interaction as a primary resource.  We
consider the scheme that is optimized for the application of quantum
gates in the network of quantum dots.  The construction of the CNOT
gate uses pulses of the exchange interaction as the only parameter
that is time dependent.  The exchange interaction itself is not
sufficient for the universal quantum computation over the ST-qubits
due to its high symmetry.  The additional symmetry breaking is provided by nonuniform, but static
magnetic fields.  These fields describe the influence both of
magnetic fields, provided by the nearby magnets, and of the coupling
to the nuclear spins in the host material via hyperfine interaction.
Depending on the scheme used for the application of the quantum gates,
the optimal geometry is either the one in which the magnetic fields
point parallel to the axis of symmetry of the SOI or
perpendicular to it.

One major problem in the realization of two-qubit quantum gates (in
particular, we consider here the CNOT gate based on conditional phase gates), is the possibility of leakage errors where the spin states defining the logical qubit
leave the computational space.  These
errors move the state of four spins from the $4$-dimensional
computational space of two qubits into some other portion of the
$16$-dimensional Hilbert space of four spin-$1/2$ particles.  We consider two
ways of addressing this problem.  One scheme possesses the axial symmetry  
due to the fact that the SOI vector and magnetic fields are parallel.  
For this `parallel scheme' we are able to construct a perfect CNOT gate, if we are  
able to control all the available parameters. Having in mind 2D architectures, we  
also consider the CNOT gate between two qubits in the case when  
the SOI vector and magnetic fields are perpendicular to each other.  Here we cannot  
prevent the leakage out of the computational space, however, we show  
that it is suppressed by a ratio between the SOI  and Zeeman energy coming 
from a strong external magnetic field.

All our constructions assume that the controlled interactions are
switched in time by rectangular pulses. Any deviations from this form of
time dependence lead to additional corrections and affect the fidelity
of the gate. 

The paper is organzied as follows. In Sec. \ref{sec:model}, we introduce the model for the double dots and effective Hamiltonians
for field gradients and exchange and spin orbit interactions.
In Sec. \ref{sec:parallel}, we consider the parallel geometry and derive the CNOT gate via the conditional phase gates and swap gates, all based
on exchange. There we also give estimates for the switching times.
The scheme for the perpendicular geometry is then addressed in Sec. \ref{sec:perp}, and we conclude in Sec. \ref{sec:conclusion}.

\section{The Effective Model\label{sec:model}}

\begin{figure}[b]
\begin{center}
\includegraphics[width=8cm]{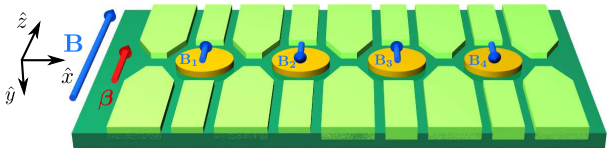}
\caption{\label{fig:setupAllZ} 
Parallel geometry: four  quantum dots (yellow discs) aligned along the x-axis in the presence of an external magnetic field $\ve B$ that is applied 
parallel to the SOI vector ${\boldsymbol \beta}$ (red arrow), which must be perpendicular to the line of the dots and which we take to be the $z$-direction.
At each dot, there is a local magnetic field $\ve B_i$ (blue arrows), also assumed to be parallel to $\ve B$,
but with alternating orientations as indicated. 
The direction of $\ve B$ defines the spin quantization axis. The dots are defined electrostatically by metallic gates (light green structures).
Each dot contains a spin-1/2, and the exchange  ($J_{ij}$) and the SOI-induced (${\boldsymbol \beta}_{ij}$) interactions  between the spins can be controlled by changing the electrostatic potential between the dots.
 The  dots 
  1 and 2 (from left to right) define the first ST-qubit, and the dots 3 and 4 the second
  ST-qubit. 
}
\end{center}
\end{figure}

\begin{figure}[t]
\begin{center}
\includegraphics[width=7.5cm]{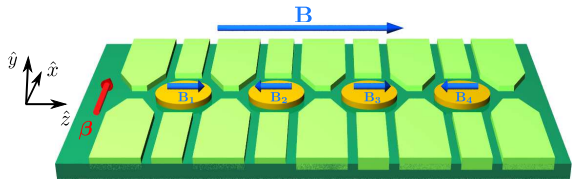}
\caption{\label{fig:setup} Perpendicular geometry: similar to the setup shown in Fig. \ref{fig:setupAllZ} but with the  difference that here the SOI vector $\boldsymbol \beta$ (red arrow) is {\it perpendicular} to the magnetic fields $\ve B$ and $\ve B_i$, which define the spin quantization axis $z$.}
\end{center}
\end{figure}

 We consider in the following singlet-triplet (ST) qubits that are implemented by two electrons confined
to a double quantum dot system  \cite{B01, L02}, see Figs. \ref{fig:setupAllZ} and ~\ref{fig:setup}. Such ST-qubits
have been realized successfully in several labs \cite{HKP+07}, and  single and two-qubit operations
have also been demonstrated recently \cite{PJT+05,BMM+10,BFN+11, Yacoby_ST}. There are several schemes for the fundamental 
 CNOT gate, which can be divided into two classes, schemes which make use of exchange interaction
 and schemes which do not, but instead rely on coupling of dipole moments \cite{Yacoby_ST}. The latter schemes has the
 disadvantage to be rather slow and also to be affected by charge noise rather strongly. Here, we focus
 on exchange-based schemes specifically adapted to quantum dots in III-V semiconducting materials,
 that have the SOI, such as in GaAs or InAs quantum dots. Although the SOI 
 is typically small compared to the level spacing of the dots, it needs to be taken into account in order to achieve
 high fidelities in gate operations.

We focus now on two such ST-qubits and assume them to be encoded in four
quantum dots that are arranged in a row, see Figs. \ref{fig:setupAllZ} and ~\ref{fig:setup}. The external magnetic
field $\ve {B}$ is assumed to give the largest energy scale and
determines the spin quantization axis $z$. The Hilbert space
of four spins-$1/2$ is spanned by $2^4=16$ basis states.  The total
spin of the system is given by
$\hat{\ve{S}}=\sum_{i=1}^{4}\hat{\ve{S}}_i$, where $\hat{\ve{S}}_i$ is a
spin-$1/2$ operator acting on the spin in a dot $i=1,2,3,4$. 
Due to axial symmetry, the z-component, $\hat{{S}}^z$, becomes a good quantum number, and the eigenstates
corresponding to $S^z=0$ span a six-dimensional subspace. We define
the computational basis of the two ST-qubits in this subspace as
\begin{align}
|00\rangle = \mid\downarrow\uparrow\downarrow\uparrow\rangle, \ \ \ |11 \rangle & = \mid\uparrow\downarrow\uparrow\downarrow\rangle,\nonumber
\\
|01\rangle = \mid\downarrow\uparrow\uparrow\downarrow\rangle, \ \ \ |10\rangle &= \mid\uparrow\downarrow\downarrow\uparrow\rangle,\label{eq:00def} 
\end{align}
where $\{0, 1\}^{\otimes 2}$ denotes the ST-qubit space, 
 `$\uparrow$' and `$\downarrow$' denote states of the quantum dot spins corresponding to the projection
$S^z_i=\pm 1/2$ on the $z$ axis. The remaining two states,
\begin{align}
|l_1\rangle = \ket{\uparrow\uparrow\downarrow\downarrow},\ \ \ |l_2 \rangle = \ket{\downarrow\downarrow\uparrow\uparrow }\label{eq:l2},
\end{align}
belong to the non-computational 
leakage space.

Besides the externally applied magnetic field $\ve B$, we allow also for  local magnetic fields $\ve B_i$ that are constant in time (at least over the switching time of the gate). Such local fields
 can be generated  e.g. by  nearby micromagnets \cite{BSO+11} or by the hyperfine field \cite{CL04,KLG03,MER02,CWD09,LYS10,FBM+09} produced by the nuclear spins of the host material. In the latter case, in order to reach a high fidelity, it is important to
perform a nuclear state narrowing \cite{CL04}, {\it i.e.} to suppress the natural variance $\delta B_i \sim A/\sqrt{N}\sim 10
\rm mT$ to a smaller value.  
 In the ideal case, one should try to
reach a limit where $|\ve {B}_i| \sim 50 \rm mT \ll |\ve B|$ and the fluctuations in
$\ve B_i$ are much smaller than $|\ve B_i|$ .

The magnetic fields $\ve {B}_i$ are assumed to point along the  external field $\ve B$, so that they preserve the axial symmetry of the problem. However, the  $\ve {B}_i$'s should have different values (to create field gradients between the dots),  a simple scenario being local fields of  opposite directions on neighboring dots, see Figs. \ref{fig:setupAllZ} and ~\ref{fig:setup}.
The corresponding Zeeman Hamiltonian is given by
\begin{equation}
H^B=\sum_{i=1}^{4} (b + b_i) \hat{S}_i^z,
\label{H_fields}
\end{equation}
where the effective magnetic fields are defined in terms of energy as $b = g\mu\sub{B}B $ and $b_i = g\mu_B B_i$, respectively, with $g$ being the electron g-factor and $\mu_B$ the Bohr magneton. 

The exchange interaction $J_{ij}(t)$ couples the electron spins of nearest neighbor dots $i$ and $j$
and can be controlled 
electrostatically \cite{HKP+07,PJT+05,BSO+11}.
If the tunnel barrier between the dots is high
we can treat them as independent. If  the tunnel
barrier is lowered and/or a detuning between the dots is applied, the two spins interact with each other, leading
to an effective description in terms of an
Heisenberg Hamiltonian \cite{LD98,BLD99,SRH+12},
\begin{equation}
H^{ex} = J_{12}(t) \hat{\ve{S}}_1 \cdot \hat{\ve{S}}_2 + J_{23}(t) \hat{\ve{S}}_2 \cdot \hat{\ve{S}}_3
+ J_{34}(t) \hat{\ve{S}}_3 \cdot \hat{\ve{S}}_4\, .
\label{hexchange}
\end{equation}
We assume that the magnetic field is sufficiently large compared to exchange energies, {\it i.e.} $J_{ij}\ll B$, to avoid admixture of
triplets via the SOI (see the discussion of perpendicular geometry below). 

We note that ideally it is best to switch the exchange
$J_{ij}$ by changing the corresponding inter-dot barrier-height or distance,
instead of detuning the double dot by a bias
$\epsilon$. Detuning  is harmful for two reasons.
First,  detuning can admix other unwanted
states (for example, (0,2)S, see Fig.~2 in \cite{SRH+12}). To analyze
the errors coming from detuning, one needs to go beyond the effective
spin Hamiltonian and consider the microscopic model for the double
dots which includes Rashba and Dresselhaus SOI and inhomogeneous fields,
see Ref. \cite{SRH+12}.  
Second, the control of $J_{ij}$ via the tunnel barrier preserves the symmetry of the charge distribution in the double dot and thus, in particular, avoids the creation of dipole moments.
In contrast, such dipole moments are unavoidable for detuning, and in the regime with $dJ_{ij}/d\epsilon\neq 0$ charge noise  can enter most
efficiently the qubit space, causing gate errors and decoherence of
the ST-qubits \cite{CL04}.
Thus, symmetric
exchange switching is expected to be more favourable for achieving high gate fidelities.

Next, we account for the effects of spin orbit interaction.
Following Refs. \cite{BSD01,BL02,SBD+03} we model the SOI by a Dzyaloshinskii-Moriya (DM) term for two neighboring quantum
dots (see e.g. Eq. (1) in \cite{BL02}),
%
%
\begin{equation}
H^{SOI}_{ij}={\boldsymbol \beta_{ij}} (t) \cdot (\hat {\ve S}_i \times \hat{\ve S}_j),
\end{equation}
where the SOI vector ${\boldsymbol \beta}_{ij}(t)$ is perpendicular to the
line connecting the dots. 
 First we consider a `parallel geometry' (see Fig. \ref{fig:setupAllZ}) where the SOI vectors
 ${\boldsymbol \beta}_{ij}$ are all parallel to each other and the magnetic fields $\ve B$ and
$\ve B_j$ are assumed to be parallel to the SOI vectors.  This preserves the
axial symmetry of the spin system, and by definition, we choose the
direction of $\ve B$ to be the $z$-axis. 
This leads
to
\begin{equation}
H^{SOI}_\parallel=\sum_{i,j}{\beta_{ij}} (t) ({\hat S}_i^x{\hat S}_j^y-{\hat S}_i^y{\hat S}_j^x)\,,
\end{equation}
where the summation runs over neighboring dots $i$ and $j$.  The
strength of the SOI, $\beta_{ij}$, depends on the distance between the dots as well
as on the tunnel coupling between them.  This
allows us to assume that both $J_{ij}(t)$ and ${\beta}_{ij}(t)$ are switched
on and off simultaneously \cite{SEA92,YHE+94,BSF10b}.

We note here that both $H^{ex}$ and $H^{SOI}_\parallel$, being axially
symmetric interactions, preserve the $z$-component of the total spin $S^z$. This means that our set-up is protected
from leakage to the subspace with $S^z \neq 0$. However, it is not
protected from the leakage 
to the non-computational space given by
Eq. (\ref{eq:l2}).
By a proper design of gates this leakage can be minimized.

Alternatively, in a `perpendicular geometry' (see Fig. \ref{fig:setup}) the axis of the
quantum dots is aligned parallel to the applied magnetic field, in the
$z$-direction.  The SOI vector ${\boldsymbol \beta}_{ij}$ is perpendicular to this, and we
take it to be in x-direction. The corresponding Hamiltonian becomes
\begin{equation}
H^{SOI}_\perp=\sum_{i,j}{\beta_{ij}} (t) ({\hat S}_i^y{\hat S}_j^z-{\hat S}_i^z{\hat S}_j^y)\, .
\end{equation}
Here, the SOI vector ${\boldsymbol \beta}_{ij}$ breaks the axial symmetry of the system, and the total spin $S^z$ is no longer a good quantum number. As a consequence, leakage
into the non-computational space $S^z\neq0$ is possible, in principle. However, this coupling involves  higher-energy states and can thus be suppressed by choosing a sufficiently large
magnetic field such that $\beta/B\ll 1$.
In contrast, the SOI does not couple states within the computational space, since the matrix elements of
$H^{SOI}_\perp$ between the states $\ket{\downarrow\uparrow}$ and $\ket{\uparrow\downarrow}$
vanish. 

\section{Parallel geometry \label{sec:parallel}}
In this section we concentrate on the parallel geometry, see Fig. \ref{fig:setupAllZ}. Using the axial symmetry of the problem we are able to 
construct a sequence of gate operations that implements 
the CNOT gate \cite{NC00}, defined on the logical ST-qubits given in Eq. (\ref{eq:00def}) by $U_{CNOT}|a,b\rangle=|a, a\oplus b\rangle$,
where $a,b=0,1$. 

One important step in this construction  is
the implementation of the $\pi/4$-gate $U_{\pi/4}$  (see discussion below and \cite{NC00}). For this gate we
propose the following scheme 
consisting of 
four steps,
\begin{equation}
C_{23}\rightarrow{(\pi_{12},\ \pi_{34})}\rightarrow C_{23}\rightarrow{(\pi_{12},\ \pi_{34})},
\label{gate_consequence}
\end{equation}
where the conditional phase gate $C_{23}$  exchange-couples the dots 2 and 3 and adds a phase factor to the two ST-qubits (see below).
The swap gates $\pi_{12}$ and $\pi_{34}$ exchange spin states on the  dots $1$ and $2$ and the dots $3$ and $4$, and can be performed in parallel.

A major issue in the implementation of  $U_{\pi/4}$ 
 is
the avoidance of  leakage errors during the coupling of qubits.
To keep qubits in
the computational space, operations on  spins $2$ and $3$ must be
constructed in such a way that the resulting gate is diagonal in the basis
given by Eq. (\ref{eq:00def}).  This can be achieved in two ways. The first
approach is to use  strong  pulses 
that lead to  fast rotations around the Bloch sphere.  The
second approach is to use adiabatic pulses that  are protected from the
leakage to states with different energies \cite{adiabatic}.  However, the adiabaticity
requires a longer pulse time. In the present work we focus on the first approach.

\subsection{Conditional phase gate $C_{23}$ \label{sec_C}}

In this subsection we describe the phase gate, $C_{23}$, involving the exchange and SOI interactions only between dot 2 and 3, while dots 1 and 4 are decoupled from  dots 2 and 3, {\it i.e.}, $J_{12}=J_{34}=0$ and  $\beta_{12}=\beta_{34}=0$. In this case, the effective Hamiltonian is given by
\begin{equation}
H^{C}=H^B+H^{ex}+H^{SOI}_\parallel=H^C_1+H^C_{23}+H^C_4,
\end{equation}
where we present it in block-diagonal form. The part of the Hamiltonian $H^C_i=(b + b_i) \hat{S}_i^z$ acts only on spins located at the dot $i=1,4$. The other part of the Hamiltonian, $H^C_{23}$,  acts on spins located at the dots 2 and 3,
\begin{align}
H^C_{23}=(b + b_2) \hat{S}_2^z&+(b + b_3) \hat{S}_3^z+J_{23}\hat{\ve{S}}_2 \cdot \hat{\ve{S}}_3\nonumber\\
&+{\beta_{23}} ({\hat S}_2^x{\hat S}_3^y-{\hat S}_2^y{\hat S}_3^x).
\end{align} Here, we assume a rectangular pulse shape for the exchange and spin orbit interactions, and from now on we treat $J_{23}$ and ${\beta_{23}}$ as 
time-independent parameters. In this case, the
unitary gate $U^C$ is a simple exponential of the Hamiltonian,
\begin{equation}
U^C=e^{-i H^C T_C}=e^{-i H_1^C T_C}e^{-i H_{23}^C T_C}e^{-i H_4^C T_C}.
\end{equation}

\begin{figure}[tb]
\begin{center}
\includegraphics[width=3.5cm]{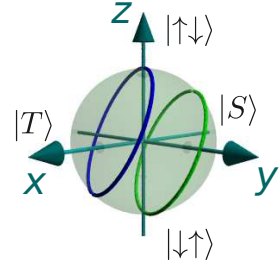}
\caption{\label{fig:23_rotation} The Bloch sphere is defined with the north pole corresponding to $\mid\uparrow\downarrow\rangle$ and the south pole corresponding to $\mid\downarrow\uparrow\rangle$. 
The effect of the unitary evolution operator $U^C_0$ [see Eq. (\ref{U_C_0})] on a state in the space $\{\mid\uparrow\downarrow\rangle, \mid\downarrow\uparrow\rangle\}$ is equivalent to the rotation on the Bloch sphere around the vector $J_{23} \ve e_x -\beta_{23} \ve e_y +  \Delta  b_{23} \ve e_z$. The conditional phase gate $C_{23}$ corresponds to the full rotation on the Bloch sphere (shown by blue and green circles).}
\end{center}
\end{figure}

The spins of dot 1 and 4 do not change in time apart from a phase factor coming from the corresponding magnetic field. In contrast, the spins of  dot 2 and 3 change in time and acquire
phases, as we describe next. For this we express $H_{23}^C$ as matrix in the basis $\{\mid\uparrow\uparrow\rangle, \mid\uparrow\downarrow\rangle, \mid\downarrow\uparrow\rangle, \mid\downarrow\downarrow\rangle\}$, 
\begin{equation}
\begin{pmatrix}
H^C_{+}& 0 & 0  \\ 
0 & H^C_{0} & 0\\
0 & 0 & H^C_{-} \\
\end{pmatrix}
\label{matrix_block},
\end{equation}
where the block-diagonal form reflects the conservation of  $S^z$. In the case of two parallel spins, the corresponding Hamiltonian is given by $H^C_{\pm}=J_{23}/4 \pm (b_2+b_3)/2$, 
which just assigns a phase to the spins. In the case of antiparallel spins, the Hamiltonian $H^C_{0}$ is given by
\begin{equation}
H^C_{0}=\frac{1}{2}\begin{pmatrix}
 -J_{23}/2 +\Delta b_{23} &  J_{23} +i \beta_{23} \\
J_{23}  -i \beta_{23}& -J_{23}/2 -\Delta b_{23} 
\end{pmatrix}
\label{matrix_with_soi},
\end{equation}
where $\Delta b_{23} =b_2 -b_3$. Here, $H^C_{0}$ describes the coupling between the states $\mid\uparrow\downarrow\rangle$ and $\mid\downarrow\uparrow\rangle$, which, in general, leads to leakage 
errors.
This leakage can be prevented by choosing the pulse duration, $T_C$, in such a way that the corresponding unitary operator $U^C_0=\exp[-i H_{0}^C T_C]$ is diagonal in the basis $\{\mid\uparrow\downarrow\rangle, \mid\downarrow\uparrow\rangle\}$. It is more convenient to consider the evolution given by $ H_{0}^C$ on the Bloch sphere. For that we rewrite $H^C_{0}$ in terms of  pseudospins,
\begin{align}
H^C_{0}&=
-J_{23}/4 + (\widetilde{J}_{23}/2)\  \ve{n}_{23}\cdot \boldsymbol{\tau},\label{H^C_0}\\
\widetilde{J}_{23}&=\sqrt{J_{23}^2 +\beta_{23}^2+(\Delta b_{23})^2},\\
\ve{n}_{23}& = \big(J_{23}, -\beta_{23}, \Delta b_{23}\big)/ \widetilde{J}_{23} \label{n_23},
\end{align}
where the unit vector $\ve {n}_{23}$
defines the rotation axis on the Bloch sphere,  see Fig. \ref{fig:23_rotation},  and the pseudospin, acting on the 
states $\{\mid\uparrow\downarrow\rangle,\mid\downarrow\uparrow\rangle\}$, is  described by the Pauli
 matrices $\vctr{\tau}$. The north pole corresponds to $\mid\uparrow\downarrow\rangle$ and the
south pole to $\mid\downarrow\uparrow\rangle$.    The exchange interaction, $J_{23}$, being
the largest scale in $H_0^C$, forces $\ve {n}_{23}$ to be aligned mostly along the $x$-axis. 
If we neglect the SOI and any  field gradients, the rotation on the Bloch sphere takes place in the $yz$-plane.
In the presence of  SOI and field gradients  the rotation axis $\ve {n}_{23}$ deviates from the $x$-axis.

The  unitary time evolution operator $U^C_0$ corresponding to $H_0^C$ takes  the form
\begin{equation}
U^C_0=\exp(i \alpha_C/2)(\cos \alpha_C+ i\ \ve{n}_{23}\cdot \boldsymbol{\tau}\sin \alpha_C ),
\label{U_C_0}
\end{equation}
where $\alpha_C=\widetilde{J}_{23}T_{C}/2$.
The duration of a pulse is determined by the condition that we obtain  full rotations on the Bloch sphere (see Fig. \ref{fig:23_rotation}),
\begin{equation}
\label{eq:t23}
T_{C} = \frac{2 \pi N_C }{\widetilde{J}_{23}},
\end{equation}
where $N_C$ is a positive integer.
Note that deviations from Eq. (\ref{eq:t23}) lead, again, to  leakage errors.

As a result, the qubit states with parallel spins on dot 2 and 3, 
acquire the phases
\begin{align}
\phi_{01} &= \frac{1}{2}(J_{23}/2-b_1+b_2+b_3-b_4) T_{C} ,\nonumber
\\ 
\label{eq:phase10}
\phi_{10} &= \frac{1}{2} (J_{23}/2+b_1-b_2-b_3+b_4) T_{C},
\end{align}
while the qubit states with antiparallel spins on dot 2  and 3  acquire the phases
\begin{align}
\phi_{11} &= \frac{1}{2} (b_1-b_4- (-1)^{N_C}J_{23}/2) T_{C}
,\nonumber
\\ 
\label{eq:phase00}
\phi_{00} &= \frac{1}{2} (-b_1+b_4- (-1)^{N_C}J_{23}/2) T_{C}.
\end{align}
Here, $\phi_{ab}$ corresponds to a phase acquired by a two-qubit state $|ab\rangle$.
We note that phases produced by the magnetic fields terms will be canceled during the second $C_{23}$
pulse after the $\pi$-pulses have been applied to the qubits.

\subsection{Swap gates $\pi_{12}$ and $\pi_{34}$ \label{sec_pi}}

In this subsection we discuss the swap gates $\pi_{12}$ and $\pi_{34}$ that exchange spin states between dot 1  and 2, and dot 3 and 4, respectively. The swap operation is a one-qubit-operation, so dots 2 and 3 should be decoupled during the swap pulse, {\it i.e.}, $J_{23}=0$ and $\beta_{23}=0$. The swap gate $\pi=(\pi_{12},\pi_{34}) $ is implemented, again, by a rectangular pulse and all parameters are assumed to stay constant during the switching process. This simplifies  the unitary evolution operator,
\begin{equation}
U^\pi=U^\pi_{12}\ U^\pi_{34}=e^{-i H_{12}^\pi T_\pi} e^{-i H_{34}^\pi T_\pi}.
\end{equation}
Further, we focus on the first ST-qubit (dots 1 and 2) and consider only $\pi_{12}$ ($\pi_{34}$ can be obtained analogously).
We also note that since $S^z$ is conserved, the final state is always given by a linear combination of the states
$\mid\nobreak\downarrow\uparrow\nobreak\rangle$ and $\mid\uparrow\downarrow\rangle$  on dot 1 and 2. Within this subspace the effective Hamiltonian is given by
\begin{equation}
H^\pi_{12}=\frac{1}{2}\begin{pmatrix}
 -J_{12}/2 +\Delta b_{12} &  J_{12} +i \beta_{12} \\
J_{12}  -i \beta_{12}& -J_{12}/2 -\Delta b_{12}\\
\end{pmatrix}\,
\label{matrix_with_soi_pi},
\end{equation}
or in pseudospin representation [see Eq. (\ref{H^C_0})],
\begin{align}
H^\pi_{12}
&=-J_{12}/4 + (\widetilde{J}_{12}/2) \  \ve{n}_{12}\cdot \boldsymbol{\tau},\label{H^pi_0}\\
\widetilde{J}_{12}&=\sqrt{J_{12}^2 +\beta_{12}^2+(\Delta b_{12})^2},\\
\ve{n}_{12}& = \big(J_{12}, -\beta_{12}, \Delta b_{12}\big)/ \widetilde{J}_{12} \label{n_23_pi},
\end{align}
where $\Delta b_{12}= b_1 -b_2$ and the Pauli matrix $\tau_i$ acts in the pseudospin space
spanned by $\mid\uparrow\downarrow\rangle$ and $\mid\downarrow\uparrow\rangle$. Again, the unit vector $\ve{n}_{12}$ defines the rotation axis. The  corresponding unitary evolution operator reduces to the form
\begin{equation}
U^\pi_{12} = \exp(i \alpha_\pi/2)(\cos \alpha_\pi+ i\ve{n}_{12}\cdot \boldsymbol{\tau} \sin \alpha_\pi),
\end{equation}
where $\alpha_\pi=\widetilde{J}_{12}T_\pi/2$.

The swap operation should exchange the states $\mid\uparrow\downarrow\rangle$
and $\mid\downarrow\uparrow\rangle$. In the absence of  SOI and field gradients,
 the unitary evolution operator $U^\pi_{12}$
corresponds to a rotation around the $x$-axis ($\ve
n_{12}=\ve e_x$) in the $yz$-plane. At  half the period, $T^0_\pi = \pi/ 2
\widetilde{J}_{12}$, a state evolves from the north to the south
pole and vice versa, ${\it i.e.}$, $U^\pi_{12}|_{T^0_\pi} \propto \tau_x$. 

However, in the presence of  SOI and/or field gradients, the rotation axis $\ve n_{12}$ is not aligned with
$\ve e_x$ (compare with Fig. \ref{fig:23_rotation}), so the trajectory
starting at the north (south) pole would never go exactly through the
south (north) pole. The corresponding deviations lead to errors in the
$\pi_{12}$ gate on the order of $\sqrt{\Delta b_{12}^2 +
  \beta_{12}^2}/J_{12}$. This means that it is impossible to produce a
perfect swap operation with only one single rectangular pulse.  
However, by applying a sequence of several rectangular pulses,
 it is still possible to produce a perfect $\pi_{12}$ gate, as we demonstrate next.

\begin{figure}[t]
\begin{center}
\includegraphics[width=8cm]{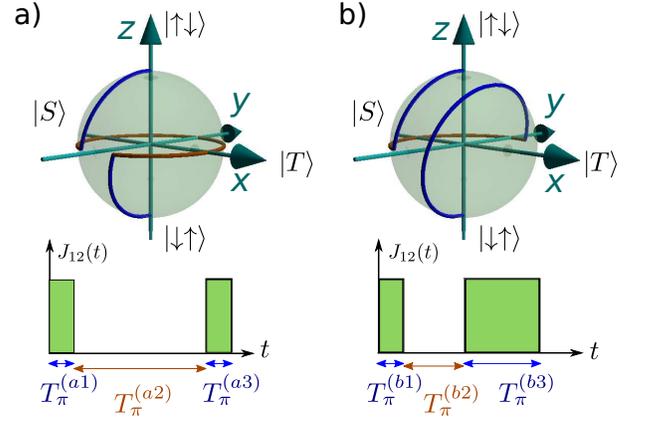}
\caption{\label{fig:Bloch_without_SOI}
Two  schemes for the swap gate, $\pi_{12}$: $\mid\uparrow\downarrow\rangle \rightarrow \mid\downarrow\uparrow\rangle$, composed of three consecutive rotations on the Bloch sphere where the offset induced by $\Delta b_{12}$ and $\beta_{12}$ is fully compensated. a)  First, starting from the north pole, we turn on $ J_{12}$ and  $\beta_{12}$ to induce rotation around $J_{12} \ve e_x -\beta_{12} \ve e_y +  \Delta  b_{12} \ve e_z$ (upper  blue arc) until we reach the equator where we turn off $ J_{12}$ and  $\beta_{12}$. Second,  we let the state precess  around the $z$-axis in the equatorial plane  until the mirror point of the starting point on the equator is reached (brown arc). Third, we induce  once more rotation around $J_{12} \ve e_x -\beta_{12} \ve e_y +  \Delta  b_{12} \ve e_z$ (lower blue arc) until we reach the south pole. Lower panel: associated rectangular gate pulses and switching times  $T_\pi^{(a1,a2,a3)}$ for the three steps.
b) Alternative scheme where during the second step the  
state precesses along the equator until it reaches the 
point diametrically opposite
to its starting point (brown arc).}
\end{center}
\end{figure}

Indeed, this goal can be achieved by the following three steps (see also Fig. \ref{fig:Bloch_without_SOI}). First, we switch on the exchange interaction $J_{12}$ between the dots 1 and 2 (this also automatically switches on the SOI $\beta_{12}$). We assume $J_{12}$ to be
larger than $\Delta b_{12}$ and/or $\beta_{12}$. The rotation axis $\ve n_{12}$ in polar coordinates is given by
\begin{align}
\ve n_{12} = (\sin \vartheta \cos\varphi, \sin \vartheta \sin \varphi, \cos \vartheta).
\end{align}
From now on we work in the coordinate system in which the $x$-axis points along $J_{12}  \ve e_x- \beta_{12} \ve e_y$, so $\varphi=0$ and 
$0\leq \cos \vartheta=\Delta b_{12}/ \widetilde{J}_{12}\leq 1/\sqrt{2}$.
The strength and the duration of the rectangular
pulse is chosen in such a way that the rotation reaches the equator of the Bloch sphere.
The initial and final vectors on the Bloch sphere are given
by
\begin{align}
\phi_{i}&=(0,0,1),\\
\phi_{f}&=(\cot \vartheta, \sqrt{\sin (2 \vartheta - \pi/2)}/ \sin \vartheta  ,0),
\end{align}
allowing us to find the rotation angle $\alpha_{\pi}^{(a1)}=
  \pi/2+\arcsin (\cot^2 \vartheta)$,
and the corresponding pulse duration
\begin{equation}
 T_\pi^{(a1)} = [\pi+2\arcsin
    (\cot^2 \vartheta)]/2 \widetilde{J}_{12}.
\end{equation}

Second, after switching off exchange and spin orbit interactions, $J_{12}=0$ and $\beta_{12}=0$,  the rotation takes place around
the $z$-axis in the equatorial plane, at a precession frequency  determined by the field gradient $\Delta b_{12}$.  The rotation
angle becomes $\alpha^{(a2)}_{\pi}=2[\pi-\arccos (\cot\vartheta)]$,
and the pulse duration is given by
\begin{eqnarray}
T_\pi^{(a2)} = \alpha^{(a2)}_{\pi}/\Delta b_{12}.
\end{eqnarray}

Finally, we repeat the first step by applying a pulse of the same strength $J_{12}$  ($\beta_{12}$) and  during the same time,
$ T_\pi^{(a3)}= T_\pi^{(a1)}$. 

An alternative scheme (b) is presented in Fig. \ref{fig:Bloch_without_SOI}b. During the second step the state evolves on the Bloch sphere only over half of the  equator, $\alpha^{(b2)}_{\pi}=\pi$, with the corresponding pulse duration  $T_\pi^{(b2)} = \alpha^{(b2)}_{\pi}/\Delta b_{12}$. The duration of the third pulse is given by
\begin{equation}
T_\pi^{(b3)}= [3\pi-2\arcsin (\cot^2
  \vartheta)]/2\widetilde{J}_{12}.
\end{equation}
The second step is the slowest one in these schemes, so the scheme
(b) has an advantage over the scheme (a) by being faster as it requires less rotation on the equator. However, scheme (b)
requires better control of parameters, since $T_\pi^{(b1)}\neq T_\pi^{(b3)}$.

Here we note that it is also possible to switch off the exchange coupling $J_{12}$ not
only on the equator but also at any other point on the Bloch sphere.  Moving away from the
equator speeds up the gate performance, but demands greater
precision in the tuning, since the rotation proceeds along a smaller
arc and in shorter time.

The scheme presented above confirms that it is possible to construct a perfect $\pi$ swap gate even in the presence of
 $z$-component of the SOI vector,
${\boldsymbol \beta}_{12}$, or local field gradients, $\Delta \ve b_{12}$, by adjusting the pulse
durations.  The other two components of the  SOI
vector couple states of different total spin  $S^z$, and thus cause leakage errors.
Therefore, it is optimal to orient the magnetic fields (defining the spin quantization axis $z$) along the SOI vector, ${\boldsymbol \beta}_{12}$.


\subsection{CNOT gate}
\begin{figure}[tb]
\begin{center}
\includegraphics[width=8cm]{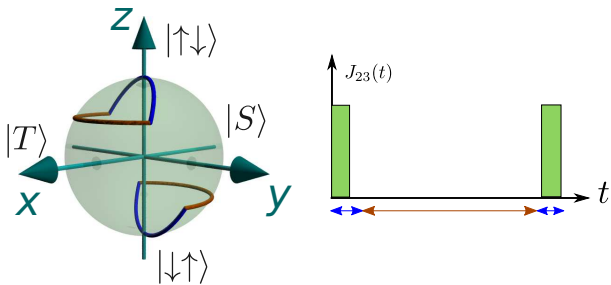}
\caption{\label{fig:23_rotation_alt} An alternative scheme for the conditional phase gate $C_{23}$ (see also Fig. \ref{fig:23_rotation}). Instead of the rotation with the pulse defined by Eqs.  (\ref{eq:t23}) and (\ref{t_231}) we consider a sequence of three pulses. During the first and the third pulses (blue arcs) the state precesses quickly acquiring the $\pi/4$-phase, see Eq. (\ref{t_231}). During the second pulse, $J_{23}=0$, the state precesses around the $z$-axis over a shorter path than the one in Fig. \ref{fig:23_rotation}. As a result, the switching  is faster.}
\end{center}
\end{figure}

After the execution of the four-step sequence given by Eq. (\ref{gate_consequence}) an initial qubit state is restored but with a phase factor  [see Eqs. (\ref{eq:phase10}) and (\ref{eq:phase00})]:
\begin{widetext}
\begin{align}
 &[00]\xrightarrow{C_{23}} e^{i \phi_{00}} [00]\xrightarrow{(\pi_{12},\ \pi_{34})}e^{i \phi_{00}}[11]\xrightarrow{C_{23}} e^{ i (\phi_{11}+\phi_{00})}[11]\xrightarrow{(\pi_{12},\ \pi_{34})}[00]e^{i (\phi_{11}+\phi_{00})},\nonumber\\
 &[11]\xrightarrow{C_{23}}  e^{i \phi_{11}} [11]\xrightarrow{(\pi_{12},\ \pi_{34})}e^{i \phi_{11}}[00]\xrightarrow{C_{23}}  e^{i (\phi_{11}+\phi_{00})}[00]\xrightarrow{(\pi_{12},\ \pi_{34})}[11]e^{i (\phi_{11}+\phi_{00})},\nonumber\\
 &[01]\xrightarrow{C_{23}}  e^{i \phi_{01}} [01]\xrightarrow{(\pi_{12},\ \pi_{34})}e^{i \phi_{01}}[11]\xrightarrow{C_{23}}  e^{ i (\phi_{01}+\phi_{10})}[11]\xrightarrow{(\pi_{12},\ \pi_{34})}[00]e^{i (\phi_{01}+\phi_{10})},\nonumber\\
&[10]\xrightarrow{C_{23}}  e^{i \phi_{10}} [10]\xrightarrow{(\pi_{12},\ \pi_{34})}e^{i \phi_{10}}[01]\xrightarrow{C_{23}}  e^{  i (\phi_{01}+\phi_{10})}[01]\xrightarrow{(\pi_{12},\ \pi_{34})}[10]e^{i (\phi_{01}+\phi_{10})},
\label{phase}
\end{align} 
\end{widetext}
with 
$\phi_{01}+\phi_{10}=-(\phi_{11}+\phi_{00})=J_{23} T_{C}/2$,
where we assume $N_C$ [see Eq. (\ref{eq:t23})] to be even. The total gate acting on the qubits as defined by Eq. (\ref{phase})  can be written in the compact form
\begin{equation}
e^{-i (J_{23} T_{C}/2) \sigma_1^z  \sigma_2^z},
\end{equation}
where the Pauli matrix $\sigma_1^j$ acts on the first ST-qubit 
(formed by dot 1 and 2) and   $\sigma_2^j$  on the second one (formed by
dot 3 and 4) with $j=x,y,z$. Choosing
\begin{equation}
T_{C} =\left( 4\pi m  -\frac{\pi}{2} \right)/J_{23},
\label{t_231}
\end{equation}
where $m$ is a positive integer,
we obtain the $\pi/4$-gate, 
\begin{equation}
U_{\pi/4}=e^{i\frac{ \pi}{4} \sigma_1^z  \sigma_2^z}.
\end{equation}
Both Eqs. (\ref{eq:t23}) and (\ref{t_231}) should be
satisfied simultaneously. For example, if $m=1$ and $N_C=2$, we get
\begin{eqnarray}
T_{C}&=\frac{\pi}{2}\sqrt{\frac{15}{\Delta
   b_{23}^2+\beta_{23}^2}},\label{T_C_1}\\ 
J_{23}&=\frac{7}{\sqrt{15}}\sqrt{\Delta
  b_{23}^2+\beta_{23}^2}\label{J_C_1}.
\end{eqnarray}
From this we can estimate the total switching time to perform the  
$\pi/4$-gate that is a sum of the switching times at each step  
[see Eq. (\ref{gate_consequence})]. For the scheme discussed in  
Subsecs. \ref{sec_C} and \ref{sec_pi} the slowest part is given by the swap  
gates $\pi_{12}$ and $\pi_{34}$, whose switching time is limited by  
field gradients (due to nuclear spins \cite{PJT+05,JPM+05,TKK+06,RPB10,BFN+11} 
and/or micromagnets \cite{,BKO+07,BSO+11}), $|\Delta b_{12}|=|\Delta b_{34}|\approx 10\ \rm mT$,  
which corresponds to $T_\pi^{(b2)}\approx 10\ \rm ns$. The gate can be  
faster if the rotations around the $z$-axis are performed not on the  
equator but more closely to the poles. This allows us to decrease the  
switching time of the swap gate to $2\ \rm ns$; however, this would require  
a more precise control over the pulses. The same trick can be used to decrease the switching time of the conditional phase gate $C_{23}$ (compare Figs. \ref{fig:23_rotation} and \ref{fig:23_rotation_alt}). If field gradients larger than $10\ \rm mT$ are used, the switching rates will be
proportionately larger.

Using the $\pi/4$-gate, 
we construct the controlled phase flip (CPF) gate
$U_{CPF}=diag(1,1,1,-1)$  (see footnote [13] in \cite{LD98}) as
\begin{equation}
U_{CPF}=
U_{\pi/4}  e^{-i\frac{\pi}{4}(\sigma_1^z +
  \sigma_2^z-1)}.
\end{equation}
Finally, we obtain the CNOT gate,
\begin{equation}
U_{CNOT}=\begin{pmatrix} \mathbb{I}&0\\ 0&\sigma_2^x
\end{pmatrix},
\end{equation}
by using the CPF gate and performing a basis rotation on qubit 2
(a single-qubit rotation by $\pi/2$ about the $y$-axis),
\begin{equation}
U_{CNOT}= e^{i \frac{\pi}{4} \sigma_2^y}U_{CPF}e^{-i \frac{\pi}{4}
  \sigma_2^y}.
\end{equation}
In summary, the full sequence of operations for the CNOT gate, $U_{CNOT}$, is given
by
\begin{equation}
e^{i \frac{\pi}{4}
  \sigma_2^y}[(\pi_{12}\pi_{34})C_{23}(\pi_{12}\pi_{34})C_{23}]
e^{-i\frac{\pi}{4}(\sigma_1^z + \sigma_2^z)}e^{-i \frac{\pi}{4}
  \sigma_2^y}.
\end{equation}

 We note again that this result has been derived under the assumption of rectangular pulse shapes. This is certainly an idealization,
 and in practice we expect  deviations from this shape to cause errors for the gates and to affect the gate fidelity. The study of this issue,
 being very important for practical purposes, requires a separate investigation and  is beyond the scope of this work.

\section{Perpendicular geometry \label{sec:perp}}

In the previous section we have discussed the parallel geometry for which
we were able to construct a perfect CNOT gate under the assumption that
we have a complete control over the parameters. The CNOT gate, together with single-qubit gates,
allows us to simulate any other quantum gate and its implementation is a crucial step toward the
realization of a quantum computer \cite{NC00}.  In a next step, many such elementary gates need to be
connected into a large network. In recent years,  the surface code \cite{WFH+11} has emerged as one of the most promising platforms
for this goal due to its large threshold of about $1\%$ for fault tolerant error correction \cite{WFH+11}.
This platform requires a two-dimensional (2D) geometry and  can be implemented in semiconductors of the type considered here. \cite{TDT+12}.

\begin{figure}[b]
\begin{center}
\includegraphics[width=6cm]{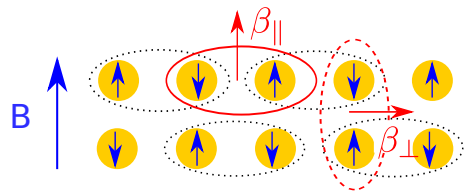}
\caption{\label{fig:2D} Schematic setup for 2D architecture. Two dots define an ST-qubit (black dotted ellipses). An external magnetic field $\ve B$ and local magnetic fields (blue arrows) are parallel. In the case of the coupling between two qubits from the same row (red solid ellipse) the SOI vector ${\boldsymbol \beta}_\parallel$ is parallel to the magnetic field $\ve B$, corresponding to the parallel geometry (see Fig. \ref{fig:setupAllZ}). In the case of  coupling between two qubits from two neighboring rows (red dashed ellipse) the SOI vector ${\boldsymbol \beta}_\perp$ is perpendicular to the magnetic field $\ve B$, corresponding to the perpendicular geometry (see Fig. \ref{fig:setup}).}
\end{center}
\end{figure}

The basic 2D scheme is illustrated in Fig. \ref{fig:2D}. There, we show an array of quantum dots  where two neighboring dots in a given row represent one ST-qubit. 
These quantum dots are embedded in a semiconductor where the SOI is of the same type in the entire structure. As a typical example we mention Rashba and/or Dresselhaus SOI that both depend on the momentum of the electron. As a result, the direction of the SOI vector ${\boldsymbol \beta}$ is always perpendicular to the line along which two quantum dots
are coupled by exchange interaction \cite{SRH+12}. Thus,
for the coupling of two such qubits in the {\it same} row, the SOI vector ${\boldsymbol \beta}$ is parallel to the external magnetic field $\ve B$, so we can use the scheme designed for the parallel geometry in the previous section. In contrast, if we want to couple two qubits from {\it neighboring} rows, we should also consider a `perpendicular geometry' (see Figs. \ref{fig:setupAllZ} and \ref{fig:2D}) in which the SOI vector ${\boldsymbol \beta}$ is perpendicular to the magnetic field $\ve B$.

This perpendicular geometry is characterized by several features. As was mentioned before, the axial symmetry in this case is broken by the SOI, leading to the coupling between  computational ($S_z=0$) and  non-computational ($S_z\neq0$) space. If the magnetic field $B$ is sufficiently large to split the triplet levels $T_{\pm}$ far away from the computational space ($g\mu_B B\gg\beta$), we can neglect this leakage. We estimate for GaAs $B=5 \rm T\approx100 \rm \mu eV/g\mu_B $. At the same time, the SOI does not couple states within the computational space, so for the realization of the phase gate $C_{23}$ and the swaps gates $\pi_{12}$ and $\pi_{34}$ we can use the same scheme as in Sec. \ref{sec:parallel} only with $\beta=0$.

\section{Conclusions \label{sec:conclusion}}

We have studied the implementation of the CNOT gate for
ST-qubits in a model that is appropriate for current experiments \cite{PJT+05,BMM+10,BFN+11}.  The setup consists
of an array of quantum dots with controlled growth direction and the
relative orientation of the dots. Pairs of neighboring dots build the
ST-qubits, where the quantization axis is determined by an externally
applied magnetic field $\ve B$.  Moreover, we introduce an
inhomogeneity in magnetic fields, $\ve B_i$, by local micromagnets or
by the hyperfine field. The resources used for time-dependent control are the
exchange interaction $J_{ij}(t)$, the field gradients $\Delta B_{ij}(t)$, and the SOI vector ${\boldsymbol \beta}_{ij} (t)$.

If the SOI vector ${\boldsymbol \beta}$ is parallel to the external  ($\ve B$) and
local magnetic fields ($\ve B_i$), we are able to construct a perfect scheme for the CNOT gate based on the
$\pi/4$-phase gate, $U_{\pi/4}$, consisting of four basic steps.  Two of the steps
involve interaction of spins that belong to different qubits, and open
the possibility of leakage errors.  Under condition of total control
over system parameters, we show that the leakage can be eliminated.
In the other two steps, the tuning of exchange interaction enables
perfect swap gates even in the presence of field gradients and SOI.

Motivated by recent results on the surface code, we shortly comment also on the 2D architecture. Here we encounter a situation in which the SOI vector ${\boldsymbol \beta}$ and the magnetic fields, $\ve B$ and $\ve B_i$, are perpendicular. In this case, the leakage to the non-computational space with $S_z\neq 0$ is inevitable. However, it can be made very small as long as $\beta/g \mu_B B \ll 1$.

Depending on the pulsing scheme, the switching times for the conditional phase gate are shown to lie in the range $1$-$100$ ns for typical GaAs parameters. Compared to the experimentally established decoherence
times of about $250$ $\mu$s \cite{BFN+11}, this switching is sufficiently fast and shows that a CNOT gate based on exchange is a promising candidate for experimental realizations.


\acknowledgments
We acknowledge discussions with Andrew Doherty, Charles Marcus, Izhar Neder, and Mark Rudner.
This work is supported by IARPA/MQCO program, DARPA/QUEST program, the Swiss NSF, 
NCCR Nanoscience, and NCCR QSIT.


\bibliographystyle{apsrev}

\end{document}